\documentclass[]{pasj01}
\usepackage{comment}
\draft

\begin{document}
\Received{}
\Accepted{}

\title{A Hard-to-Soft State Transition of Aquila X-1 Observed with Suzaku}

\author{Ko \textsc{Ono}\altaffilmark{1}%
\thanks{Example: Present Address is xxxxxxxxxx}}
\altaffiltext{1}{Department of Physics, The University of Tokyo, 7-3-1 Hongo, Bunkyo-ku, Tokyo 113-0033, Japan}
\email{ono@juno.phys.s.u-tokyo.ac.jp}

\author{Kazuo \textsc{Makishima}\altaffilmark{2}}
\altaffiltext{2}{MAXI Team, Global Research Cluster, The Institute of Physical and Chemical Research, 2-1- Hirosawa, Wako, Saitama 351-0198, Japan}

\author{Soki \textsc{Sakurai},\altaffilmark{1}}

\author{Zhongli \textsc{Zhang}\altaffilmark{3}}
\altaffiltext{3}{Shanghai Astronomical Observatory, Chinese Academy of Sciences, 200030 Shanghai, China}

\author{Kazutaka \textsc{Yamaoka}\altaffilmark{4,5}}
\altaffiltext{4}{Division of Particle and Astrophysical Science, Graduate School of Science, Nagoya University, Furo-cho, Chikusa-ku, Nagoya, Aichi 464-8602, Japan}
\altaffiltext{5}{Institute for Space-Earth Environmental Research (ISEE), Nagoya University, Furo-cho, Chikusa-ku, Nagoya, Aichi 464-8601, Japan}

\author{Kazuhiro \textsc{Nakazawa}\altaffilmark{1}}


\maketitle

\begin{abstract}
The recurrent soft X-ray transient Aquila X-1 was observed with Suzaku for a gross duration of 79.9~ks, on 2011 October 21 when the object was in a rising phase of an outburst. During the observation, the source exhibited a clear spectral transition from the hard state to the soft state, on a time scale of $\sim$30~ks. Across the transition, the 0.8--10~keV XIS count rate increased by a factor $\sim$3, that of HXD-PIN in 15-60~keV decreased by a similar factor, and the unabsorbed 0.1--100~keV luminosity increased from 3.5 $\times$10$^{37}$ erg s$^{-1}$ to 5.1 $\times$10$^{37}$ erg s$^{-1}$. The broadband spectral shape changed continuously, from a power-law like one with a high-energy cutoff to a more convex one. Throughout the transition, the 0.8--60~keV spectra were successfully described with a model consisting of a multi-color blackbody and a Comptonized blackbody, which are considered to arise from a standard accretion disk and a closer vicinity of the neutron star, respectively. All the model parameters were confirmed to change continuously, from those typical in the hard state to those typical of the soft state. More specifically, the inner disk radius decreased from 31~km to 18~km, the effects of Comptonization on the blackbody photons weakened, and the electron temperature of Comptonization decreased from 10~keV to 3~keV. The derived parameters imply that the Comptonizing corona shrinks towards the final soft state, and/or the radial infall velocity of the corona decreases. These results reinforce the view that the soft and hard states of Aql X-1 (and of similar objects) are described by the same ``disk plus Comptonized blackbody'' model, but with considerably different parameters.
\end{abstract}

\section{Introduction}\label{sec:introduction}

It is well known that black--hole binaries make transitions among several distinct spectral states, mainly depending on their luminosity. Among these transitions, the best recognized are those between the soft and the hard states (e.g., \cite{Remillard06}). Similarly, Neutron--Star (NS) Low--Mass X--ray Binaries (LMXBs), consisting of mass-donating low--mass stars and mass-accreting weakly-magnetized NSs, exhibit transitions between their hard and soft states, which are realized when their luminosities are lower and higher than $\sim 4\%$ of their Eddington value, respectively (e.g., \cite{Asai12}).
\par
Although the two states of LMXBs were often studied independently, broad-band Suzaku data have allowed us to explain the LMXB spectra in both these two states using essentially the same three ingredients \citep{Sakurai12,Sakurai14,Zhang14,Ono16}: a multi-color disk (MCD) emission from an optically-thick accretion disk, a blackbody (BB) component from the NS surface, and a hot electron cloud (a ``corona'') which Comptonizes the BB (or sometimes BB+MCD) photons. In terms of this spectral composition, the two states are distinguished primarily by a marked difference in the Comptonization strength (much stronger in the hard state). Furthermore, the MCD and BB components both have generally larger radii and lower temperatures in the hard state.
\par
In order to further strengthen the above unified view of LMXBs, we can utilize their spectral state transitions. That is, the identification of the spectral ingredients and the spectral decompositions can be made less ambiguous, by continuously tracing whether each spectral component in one state really transfers into the corresponding one in the other state; thus, we will be able to connect the spectral modeling in the less-understood hard state to those in the better-understood soft state. As a result, such attempts have been carried out by many authors \citep{Dai10,Egron13,Seifina15}. Although the studies were often hampered by difficulty to catch such transitions which are rare and generally difficult to predict with a sufficient accuracy \citep{Dai10,Egron13}, some authors successfully detected transitions (e.g., \cite{Church14,Lin07}) utilizing, for example, RXTE, and found continuous changes of the spectral parameters. In the present study, we attempt to update these results using improved instruments.
\par
For the above purpose, the Suzaku observatory, with its broad energy coverage, is particularly suited. We hence searched the Suzaku archive for a suitable data set, and found an ideal one. Namely, a Suzaku observation of Aquila X-1, made on 2011 October 21 for a gross duration of 79.9~ks in a rising phase of an outburst, very fortunately caught a hard-to-soft spectral transition of this typical recurrent LMXB. Aquila X-1 (hereafter Aql X-1) repeats outbursts, in which the 2--10 keV flux increases by a factor of $>200$ \citep{Sakurai12}, and is associated with an optical counterpart with a magnitude of $V=$15--$19$ \citep{Gottwald91}. It exhibits type I bursts, by which the distance has been constrained to be 4.4--5.9 kpc assuming that the burst peak luminosity reaches the Eddington luminosity for an assumed NS mass of 1.4 $M_{\odot}$ \citep{Jonker04}. In the present paper, we analyzed this precious data set, employing a distance of $D=5.2$~kpc \citep{Sakurai12}.



\section{Observation and Data Reduction}\label{sec:obs_and_data}

\begin{figure}[htbp]
 \begin{center}
	\FigureFile(170mm,100mm){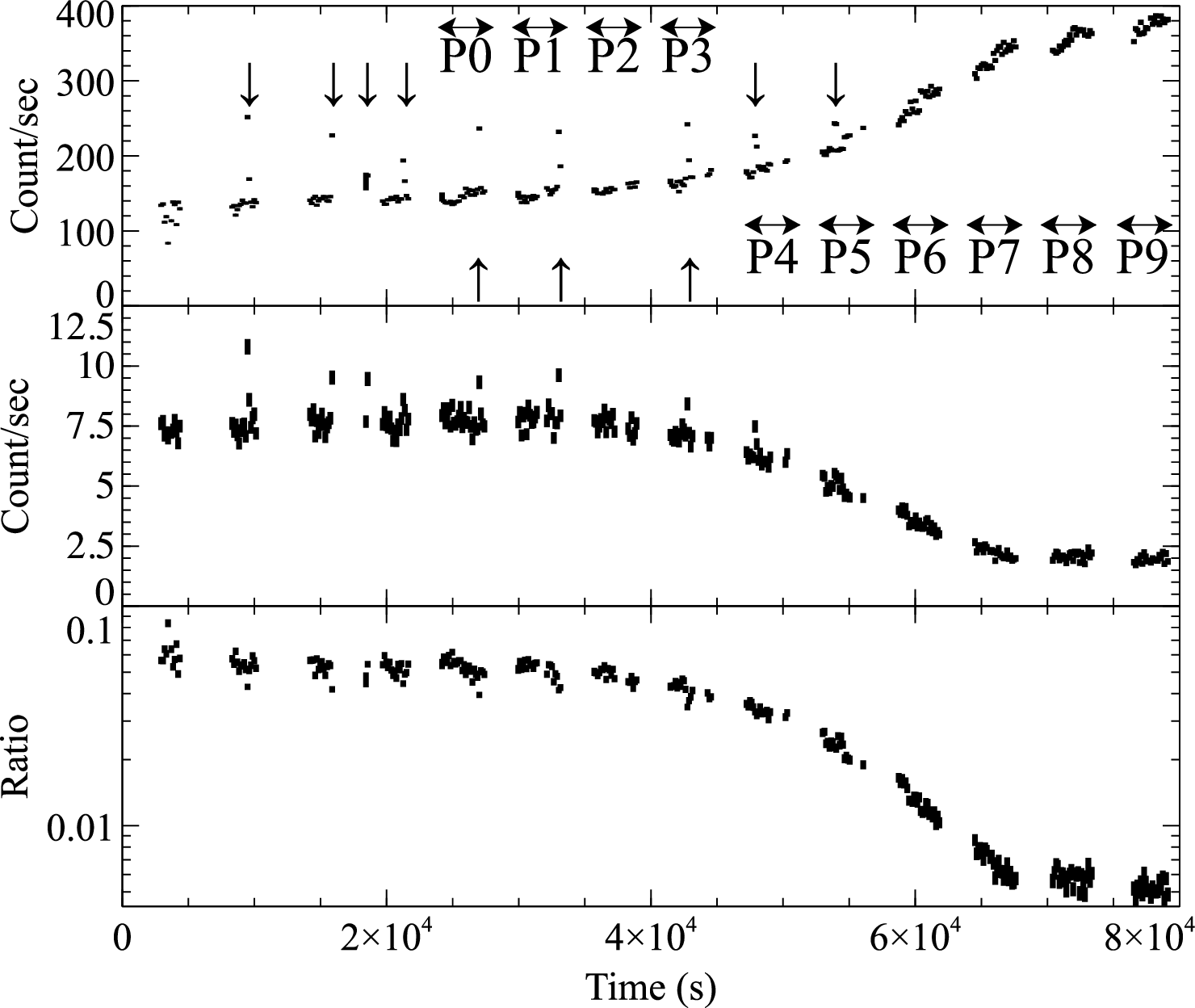}
 \end{center}
\caption{Background--subtracted light curves of Aql X-1 with 128~s binning obtained with (a) XIS (XIS 0 plus XIS 3) in 0.8--10~keV and (b) HXD-PIN in 15--60~keV. Panel (c) is the HXD--PIN vsXIS hardness ratio. Arrows in panel (a) indicate Type I X-ray bursts. Time 0 corresponds to 2011 October 21 12:51:33}\label{fig:lightcurve}
\end{figure}

\subsection{Observation}\label{sec:observation}
The data set of Aql X-1 to be analyzed here (ObsID 406010020) was acquired from 2011 October 21 12:51:33 UT for a gross duration of 79.9 ks, with the XIS and the HXD onboard. The source was placed at the ``XIS nominal'' position. To reduce pile--up effects, The XIS was operated in the 1/4 window mode with a read--out time of 2.0~s. Furthermore, the burst mode option was employed, in which the data are accumulated only for 0.5~s per the 2.0~s read-out time.

\subsection{XIS data reduction}\label{sec:xis_data_reduction}
We utilized the XIS 0 and XIS 3 events of GRADE 0, 2, 3, 4 and 6. On--source events were accumulated in a circle with a radius of $2'.5$ around the image centroid. To further reduce the pile--up effects down to $<3\%$ \citep{Yamada12}, events in a circle with a radius of $1'$ at the center were discarded. Background events to be subtracted from the on--source data were taken from an annular region with the inner radius of $4'$ and the outer radius of $5'$.
\par
Figure \ref{fig:lightcurve} (a) presents the background--subtracted 0.8--10~keV XIS (XIS 0 plus XIS 3) light curve derived in this way. It shows 9 Type I bursts, while two more were lost after we adjusted the XIS and HXD exposures. Thus, the source clearly brightened up by a factor of $\sim$3 in the XIS band, mainly across $\sim30$~ks from $\sim40$~ks to $70$~ks after the observation started.
\par
As shown in figure \ref{fig:lightcurve}, we define 10 data segments, to be named P0, P1, $\cdots$ to P9, each lasting for $\sim 2$~ks but with the Type I bursts excluded. The net XIS exposure summed over P0--P9 is 6.4 ks, which is much shorter than that with the HXD (see section \ref{sec:hxd_data_reduction}) because of the burst option (section \ref{sec:observation}).

\subsection{HXD data reduction}\label{sec:hxd_data_reduction}
Excluding the Type I bursts, HXD cleaned events were accumulated over the same ten periods, P0--P9, with a summed net exposure of 22.4 ks after dead time correction. We used non-X-ray background (NXB) events distributed by the HXD team, to reproduce NXB spectra \citep{Fukazawa09} which are subtracted from the on--source data. Cosmic-X-ray background (CXB) included in the on--source data was accounted for by adding a fixed CXB model to the spectral models describing the Aql X-1 emission (see section \ref{sec:fit1}). Figure \ref{fig:lightcurve} (b) shows the dead--time--corrected 15--60~keV light curve thus derived with HXD--PIN. There, the NXB contribution, estimated as $\lesssim0.7$~cts s$^{-1}$, was already subtracted. Thus, the HXD--PIN count rate decreased clearly, by a factor of 3--4, in anti--correlation with that of XIS.
\par
Figure \ref{fig:lightcurve} (c) shows the ratio of the HXD-PIN to XIS count rates. It clearly shows a continuous and monotonic decrease, which we take as a convincing signature of a hard--to--soft state transition. The associated spectral change is the most significant across P4--P7. Thus, the present observation fortunately witnessed a very rare occasion of the state transition. For reference, the present episode on MJD 55855, is seen in the long-term MAXI and Swift light curves of figure 3 of \citet{Asai15}, as a rapid rise in the 2--10~keV MAXI/GSC intensity and a drop in the 15--50~keV Swift/BAT count rate.
\par
The source was detected significantly with HXD-GSO over P0--P4, but not in P5--P9. In analyzing the P0--P4 spectra, we hence utilize the HXD--GSO data up to an energy where the signal becomes comparable to that of the systematic NXB uncertainty \citep{Fukazawa09}.

\section{Spectral Analysis}

\begin{figure}[htbp]
	\FigureFile(80mm,80mm){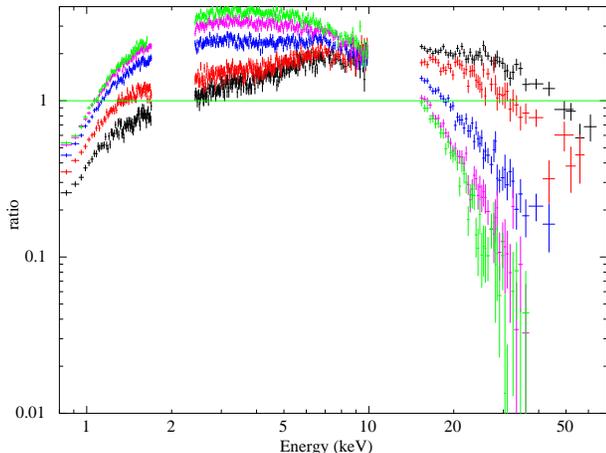}
\caption{ Background--subtracted XIS 0 plus XIS 3 and HXD spectra of Aql X-1, taken in the time intervals of P0 (black), P4 (red), P6 (blue), P7 (magenta), and P9 (green). They are all normalized by a common power-law, with a photon index of 2.0 and normalization of 1 ph cm$^{-2}$ s $^{-1}$ keV$^{-1}$ at 1~keV.}\label{fig:spectra}
\end{figure}

Figure \ref{fig:spectra} shows XIS+HXD spectra in several time intervals, all normalized to a common power law model with a photon index of 2. The P0 and P4 spectra, with a power law like shape (somewhat harder than the normalizing power-law) and a cut off around 30--40~keV, exhibit characteristics which are typical of the hard state. In contrast, the P6, P7 and P9 spectra show those of the soft state. Therefore, we reconfirm our inference made in section \ref{sec:obs_and_data}, that the observation caught a hard--to--soft state transition. The spectra have a pivot at $\sim$8~keV; the flux density at $\sim$30~keV decreased nearly by an order of magnitude, while that at 3~keV tripled.

\subsection{Continuum modeling}\label{sec:fit1}
To the ten spectra, we applied the model consisting of MCD and Comptonized BB emission using XSPEC (version 12.9.0). As the MCD component, an XSPEC model {\tt diskbb} was utilized.  An XSPEC model {\tt nthcomp} \citep{Zdziarski96,Zycki99} was utilized as the Comptonized BB emission, because it is suitable for Comptonization with optical depths $\tau>2$ \citep{Zdziarski96} which is the case in the soft state \citep{Sakurai12} and presumably at around the transition \citep{Ono16}. The seed photon source was chosen to be the BB arising from the NS surface. The absorption was taken into account by a multiplicative XSPEC model {\tt wabs}, with the column density fixed at $3.6\times 10^{21}{\rm cm^{-2}}$ which was obtained independently by \citet{Sakurai14} and \citet{Gatuzz16}. Thus, the model we used is expressed as {\tt wabs$\times$(diskbb+nthcomp)}. The free parameters are the inner disk temperature $T_{\rm in}$, the MCD normalization corresponding to the inner disk radius $R_{\rm in}$, the BB temperature $T_{\rm BB}$, the BB radius (assuming a spherical geometry) $R_{\rm BB}$, the Comptonizing electron temperature $T_{\rm e}$, and its optical depth $\tau$.
\par
To the continuum model as constructed above, we added an analytical CXB model \citep{Boldt87} expressed as\par
${\rm CXB}(E)=9.41 \times 10^{-3}(\frac{E}{1 \rm keV})^{-1.29}{\rm exp}(-\frac{E}{40 \rm keV})$,
\\where the units are ${\rm photons\ cm^{-2}\ s^{-1} keV^{-1} FOV^{-1}}$, and $E$ is the energy in keV.
 The energy range of 1.7--2.2 keV was ignored to avoid calibration uncertainties associated with the silicon K-edge and gold M-edge. The XIS vs. HXD cross normalization was expressed by multiplying the model for the HXD with the constant factor 1.16 \citep{Kokubun07}.
 \par
The fit was good with $\chi^{2}_{\nu}\sim1.1$ for all the 10 spectra. The fit results for P0, 4, 6, 7 and 9 are shown in figure \ref{fig:residuals1}, together with residuals of all periods. As expected, the softer and harder parts of the continuum were reproduced by the {\tt diskbb} and {\tt nthcomp} components, respectively. In addition, the MCD component clearly became more dominant from P0 to P9, while the Comptonization component became weaker with decreasing cutoff energy, indicating a decrease in $kT_{\rm e}$. The obtained model parameters are summarized in table \ref{table:parameters1} (after incorporating the iron line; subsection \ref{sec:fit2}), and plotted in figure \ref{fig:parameters_plots} against the time (panels a--d) and the luminosity (e--g). There, the values of $R_{\rm BB}$, describing the seed photon source, were calculated assuming that the Comptonization process conserves the photon number. The value of $R_{\rm in}$ was deduced from the MCD normalization $N_{\rm MCD}$ as\par
$R_{\rm in} = \xi \kappa ^2 \left(N_{\rm MCD}\right)^{\frac{1}{2}}\left(\frac{D}{10~{\rm kpc}}\right)\left(\cos i \right)^{-\frac{1}{2}}$
\\where $D=5.2$~kpc is the employed distance, $i=45^{\circ}$ is an inclination assumed following \citet{Sakurai12}, while $\xi=0.412$ and $\kappa=1.7$ describe effects of the inner boundary condition of the disk, and the color-hardening, respectively \citep{Kubota98}. Panel (d) in figure \ref{fig:parameters_plots} and table 2 show luminosities of the individual emission components. Here, the seed BB luminosity $L_{\rm BB}$ was calculated from the BB temperature and the 0.1--50 keV photon flux of {\tt nthcomp}, assuming spherical emission. The disk luminosity was obtained by the 0.1--10 keV energy flux of {\tt diskbb} without correction for the inclination. Finally, the luminosity carried by the Comptonization process, $L_{\rm comp}$, was derived by subtracting $L_{\rm BB}$ calculated above from the 0.1--100~keV luminosity of {\tt nthcomp}.
\par
Figure \ref{fig:parameters_plots} reveals several important properties of the transition. First, throughout P0--P9, we confirm $R_{\rm BB}<R_{\rm NS}\equiv 12~{\rm km}<R_{\rm in}$ and $T_{\rm in}<T_{\rm BB}<T_{\rm e}$, which makes the model physically reasonable. Second, the parameters from P0--P3 are consistent with those obtained by Sakurai et al. (2014) from Aql X-1 in its hard state, and those from P8--P9 agree with the soft-state parameters of this object \citep{Sakurai12}. Third, the parameters changed continuously and monotonically from the hard to the soft state. Fourth, $L_{\rm BB}$ keeps a constant fraction ($\sim 30\%$) of the total luminosity $L_{\rm t}$, as predicted by the virial theorem (the deviation from the rigorous prediction of 50\% is considered in section \ref{sec:budget}). Finally, the transition is mainly characterized by a monotonic decrease in $L_{\rm comp}$ and the associated increase in $L_{\rm disk}$, as reflected most directly in the increase of the XIS count rate and the decrease of the HXD signal (figure \ref{fig:lightcurve}). This agrees with the theoretical picture of the hard-to-soft transition; the inner accretion flow changes from an optically-thin/geometrically-thick corona to the optically-thick/geometrically-thin disk.
\par
Let us examine whether the transitions are reproducible from one outburst to another, putting aside the hysteresis between the rising and declining phases. Figure 14 in \citet{Lin07} shows that the hard-to-soft state transition of Aql X-1 in 2000 took place within $\sim1$ day which is consistent with our result of $\sim30$~ks ($\sim8$ hours) and with those of \citet{Asai15} (discussed in section \ref{sec:timescale} in detail). In the hard-to-soft transition of \citet{Lin07}, the total luminosity increased from $\sim3\times10^{37}$~erg s$^{-1}$ to $\sim5\times10^{37}$~erg s$^{-1}$, which coincides with the values we obtained in table \ref{table:luminosities}. Furthermore, in figure 8 of \citet{Lin07}, the ratio of the Comptonized blackbody luminosity (corresponding with $L_{\rm BB}+L_{\rm comp}$) to the thermal component luminosity (corresponding with $L_{\rm disk}$), decreased from 0.8 to 0.5, again in good agreement with table \ref{table:luminosities}. Thus, the rapid but continuous spectral changes seen over P4--P7 of the present data are considered to represent essentially the same phenomenon as the transition caught by \citet{Lin07}.
 
\subsection{Line emission}\label{sec:fit2}
\begin{figure*}[htbp]
	\FigureFile(180mm,140mm){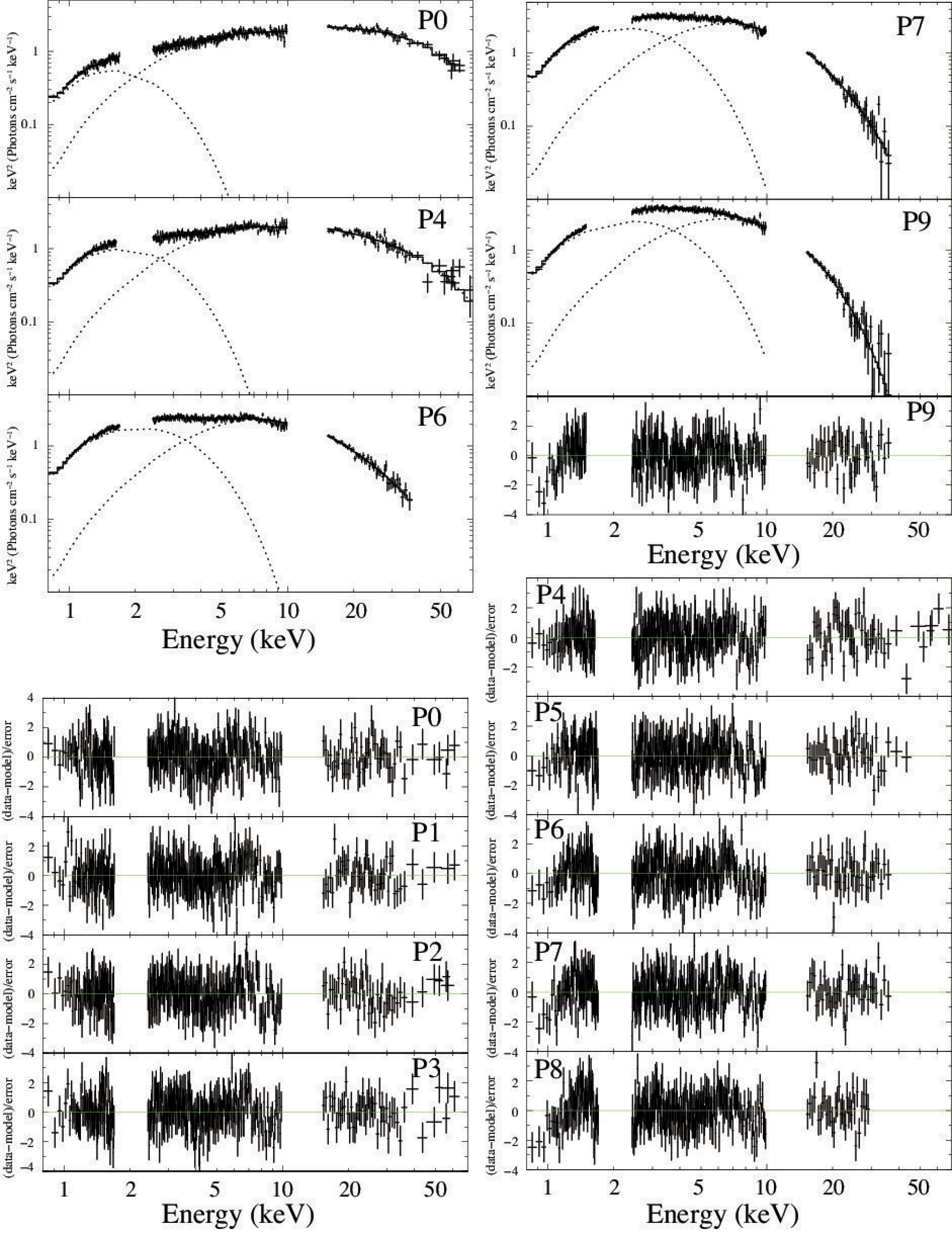}
\caption{The XIS 0 plus XIS 3 ($<10$~keV), HXD-PIN ($15-60$~keV), and HXD-GSO ($>50$~keV) spectra of P0, P4, P6, P7 and P9, fitted with {\tt diskbb+nthcomp}. The data and the best-fit model are presented in $\nu F \nu$ forms. The fit residuals are shown for all the 10 periods.}\label{fig:residuals1}
\end{figure*}

\begin{figure*}[htbp]
 \begin{center}
	\FigureFile(178mm,178mm){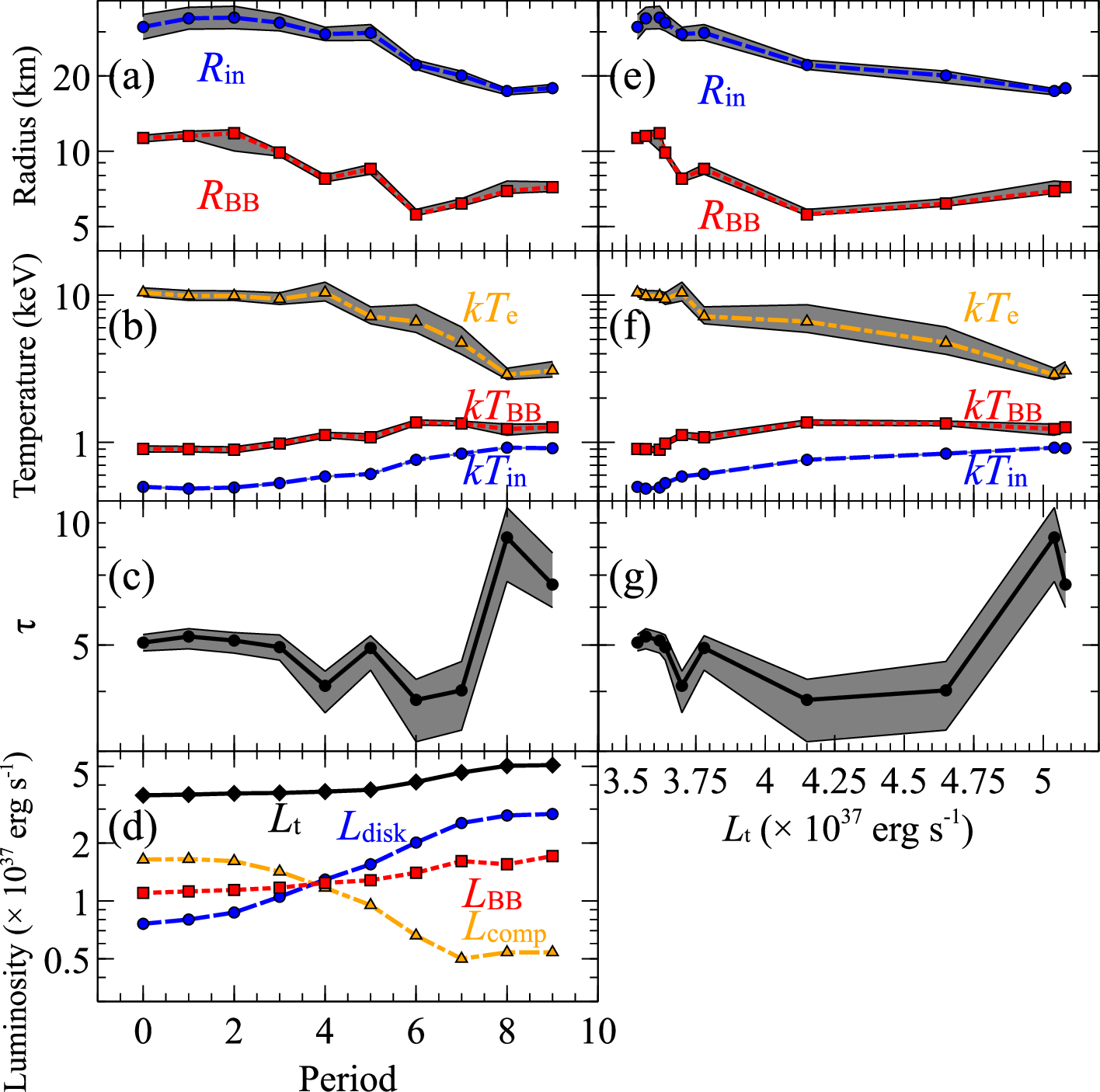}
 \end{center}
\caption{The evolution of the spectral model parameters, against the time (panels a--d) and the total luminosity (panels e--g). (a) The values of $R_{\rm in}$ (filled circles) and  $R_{\rm BB}$ (filled squares). (b) The three temperatures, (c) The optical depth of Comptonization. (d) The luminosity of each process, derived as described in text. Panels (e)--(g) the same as (a)--(c) respectively, but abscissa is taken to be the total luminosity.\label{fig:parameters_plots}}
\end{figure*}

In figure \ref{fig:residuals1}, residuals in most of the fits reveal a broad positive structure at 6.6--6.9~keV. This is considered to be an iron line, produced on the disk when illuminated by the corona, and broadened due to the relativistic effects near $R_{\rm in}$. Thus, we employed an XSPEC model {\tt diskline} \citep{Fabian89} which incorporates the special and general relativistic effects around a black hole. The spectra were fitted again with a model {\tt wabs $\times$ (diskbb+nthcomp+diskline)}, by fixing $R_{\rm in}$ of {\tt diskbb} and {\tt diskline} to the value obtained in section \ref{sec:fit1} and keeping the inclination angle at $45^\circ$. The emissivity parameter $\beta$ was between -10 and 10, and the central object mass as $1.4$~$M_{\odot}$. By introducing these 3 degrees of freedom, the fit chi--squared decreased by 11-26 in P0--P8 and by 3 in P9. The energy center was found at $\sim 6.6$~keV and did not change significantly during the observation. The equivalent width (EW) against the total continuum was $\sim100$--$200$~eV in most of the periods.

\begin{table*}[htbp]
\caption{The parameters and the 90\% statistical errors obtained by the fit with {\tt diskbb+nthcomp+diskline}.\label{table:parameters1}}
\begin{tabular}{lcccccccccc}
\hline
\hline
   & \multicolumn{2}{l}{{\tt diskbb}} & \multicolumn{4}{l}{{\tt nthcomp}} & \multicolumn{3}{l}{{\tt diskline}\footnotemark[$*$]} &  \\ \hline
	&$T_{\rm in}$~(keV)			&$R_{\rm in}$~(km)\footnotemark[$\dagger$]		&$T_{\rm BB}$~(keV)	&$R_{\rm BB}$~(km)\footnotemark[$\ddagger$]		&$T_{\rm e}$~(keV)		&$\tau$\footnotemark[$\S$]			&$E_{\rm c}$~(keV)			&$\beta	$			&EW~(eV)	&$\chi^{2}_\nu(\nu)$     \\
P0	&0.499$\pm0.003$			&$31.4\pm^{3.8}_{3.3}$		&0.90$\pm0.04$		&$11.3\pm^{0.3}_{0.4}$	&10.4$\pm^{0.8}_{0.7}$	&5.1$\pm^{0.2}_{0.2}$		&$6.6\pm^{0.4}_{0.1}$	&$-5.1\pm^{1.8}_{-}$		&170			&1.08 (333)   \\
P1	&0.485$\pm0.003$			&34.0$\pm^{3.7}_{3.2}$		&0.90$\pm0.04$		&$11.5\pm^{0.5}_{0.3}$	&9.9$\pm^{0.8}_{0.7}$	&5.2$\pm^{0.2}_{0.3}$		&$6.9\pm^{0.2}_{0.3}$	&$-2.5\pm^{0.9}_{1.3}$	&150			&1.02 (331)    \\
P2	&0.494$\pm0.003$			&34.2$\pm^{3.3}_{3.8}$		&0.89$\pm0.04$		&$11.8\pm^{0.4}_{1.8}$	&9.9$\pm^{0.8}_{0.7}$	&5.1$\pm^{0.2}_{0.3}$		&$6.6\pm0.2$			&$-3.0\pm^{0.7}_{3.8}$	&180			&0.98 (332)    \\
P3	&0.529$\pm0.003$			&32.6$\pm^{2.8}_{2.4}$		&0.98$\pm0.04$		&$9.9\pm0.3$			&9.4$\pm^{1.0}_{0.8}$	&4.9$\pm^{0.3}_{0.3}$		&$6.4\pm^{0.09}_{0.08}$	&$-6.4\pm^{2.9}_{-}$		&210			&1.03 (333)    \\
P4	&0.587$\pm^{0.002}_{0.003}$	&29.4$\pm^{1.9}_{1.7}$		&1.12$\pm0.05$		&$7.8\pm^{0.3}_{0.2}$	&10.4$\pm^{1.8}_{1.3}$	&4.1$\pm^{0.3}_{0.5}$		&$6.5\pm^{0.2}_{0.3}$	&$-2.8\pm^{1.1}_{2.2}$	&130			&1.07 (327)    \\
P5	&0.610$\pm0.003$			&29.7$\pm^{2.3}_{2.0}$		&1.08$\pm0.06$		&$8.5\pm0.4$			&7.2$\pm^{1.2}_{0.8}$	&4.9$\pm^{0.3}_{0.5}$		&$6.5\pm^{0.2}_{0.1}$	&$-4.8\pm^{1.7}_{-}$		&170			&1.04 (328)    \\
P6	&0.762$\pm0.002$			&22.1$\pm^{1.0}_{0.9}$		&1.37$\pm^{0.05}_{0.06}$	&$5.6\pm^{0.3}_{0.1}$	&6.6$\pm^{2.0}_{1.0}$	&3.8$\pm^{0.4}_{0.7}$		&$6.5\pm0.2$			&$-2.3\pm^{0.8}_{1.5}$	&98			&1.07 (326)    \\
P7	&0.839$\pm^{0.002}_{0.003}$	&20.1$\pm^{0.8}_{1.4}$		&1.34$\pm0.05$		&$6.2\pm^{0.3}_{0.2}$	&4.8$\pm^{1.3}_{0.8}$	&4.0$\pm^{0.6}_{0.7}$		&$6.6\pm^{0.2}_{0.4}$	&$-3.5\pm^{1.2}_{-}$		&110			&1.09 (326)    \\
P8	&0.921$\pm^{0.004}_{0.005}$	&17.4$\pm^{0.4}_{0.6}$		&1.23$\pm^{0.10}_{0.11}$	&$6.9\pm^{0.7}_{0.2}$	&2.9$\pm^{0.3}_{0.2}$	&8.4$\pm^{1.3}_{1.6}$		&$6.6\pm^{0.2}_{0.6}$	&$-3.5\pm^{1.2}_{-}$		&99			&1.13 (318)    \\
P9	&0.913$\pm0.004$			&17.9$\pm0.6$				&1.27$\pm^{0.07}_{0.09}$	&$7.2\pm^{0.4}_{0.3}$	&3.1$\pm^{0.5}_{0.3}$	&6.7$\pm^{1.1}_{0.7}$		&$6.4\pm^{0.34}_{-}$	&$-6.4\pm^{0.3}_{0.2}$	&46			&1.18 (298)    \\ \hline
\multicolumn{4}{@{}l@{}}{\hbox to 0pt{\parbox{170mm}{\footnotesize
\par\noindent
\footnotemark[$*$] Lower limit is not shown when is not obtained.
\par\noindent
\footnotemark[$\dagger$] Inclination is taken into account multiplying the obtained value in \S \ref{sec:fit1} by $1/\cos45^\circ$. Errors are calculated from the model {\tt diskbb+nthcomp}.
\par\noindent
\footnotemark[$\ddagger$] Calculated by the photon number conservation before and after Comptonization.
\par\noindent
\footnotemark[$\S$] Errors are calculated from the error of either $\Gamma$ or $kT_{\rm e}$. Larger one is shown here between these two.
}\hss}}
\end{tabular}
\end{table*}

\begin{table}[htbp]
\caption{Luminosities of the model components and its sum. \label{table:luminosities}}
\begin{tabular}{lcccc}
\hline
\hline
	&\multicolumn{4}{l}{Luminosities\footnotemark[$*$] ($\times10^{37}$erg s$^{-1}$)} \\ \hline
P	&$L_{\rm t}$	&$L_{\rm disk}$\footnotemark[$\dagger$]	&$L_{\rm BB}$\footnotemark[$\ddagger$]		&$L_{\rm comp}$\footnotemark[$\S$]\\
0	&3.54			&0.76					&1.10			&1.64 \\
1	&3.57			&0.80					&1.12			&1.65 \\
2	&3.62			&0.87					&1.14			&1.61 \\
3	&3.64			&1.05					&1.17			&1.42 \\
4	&3.70			&1.29					&1.24			&1.17 \\
5	&3.78			&1.55					&1.28			&0.95 \\
6	&4.15			&2.09					&1.40			&0.66 \\
7	&4.65			&2.54					&1.61			&0.50 \\
8	&5.04			&2.78					&1.55			&0.54 \\
9	&5.08			&2.83					&1.71			&0.54 \\ \hline
\multicolumn{5}{@{}l@{}}{\hbox to 0pt{\parbox{85mm}{\footnotesize
\par\noindent
\footnotemark[$*$]Values are rounded off to the second decimal place. $L_{\rm t}$ is not necessarily the same as the sum of the rest three components, $L_{\rm disk}$, $L_{\rm BB}$ and $L_{\rm comp}$.
\par\noindent
\footnotemark[$\dagger$] Luminosity (0.1--10~keV) of the disk. Inclination is not considered.
\par\noindent
\footnotemark[$\ddagger$] Luminosity (0.1--100~keV) of the BB.
\par\noindent
\footnotemark[$\S$] Luminosity (0.1--100~keV) transferred from the corona to the BB photons.
}\hss}}
\end{tabular}
\end{table}

\section{Discussion}\label{sec:Discussion}
During an 80 ks of pointing with Suzaku onto Aql X-1 while the source was in the rising phase of an outburst, a complete hard-to-soft state transition was captured. Although the transition involved a drastic spectral hardening, the broad-band (typically 0.8--60 keV) spectrum was reproduced, throughout the event, with the same {\tt diskbb+nthcomp+diskline} model. The obtained parameters evolved continuously as in figure \ref{fig:parameters_plots}. As indicated by table \ref{table:luminosities}, the total luminosity increased monotonically from $3.5\times10^{37}$ erg s$^{-1}$ to $5.1\times10^{37}$ erg s$^{-1}$, which fall on the typical boundary of the hard-to-soft state transition as discussed in the last paragraph of section \ref{sec:fit1}. To be more specific, these values are even higher by a factor of 2-3 than those measured in the soft state ($1.5\times10^{37}$ erg s$^{-1}$) before a soft--to--hard state transition \citep{Sakurai14}. This inversion is considered to be a hysteresis effect of spectral transition \citep{Maccarone03,Asai12}.

\subsection{Evolution of the spectral parameters} \label{sec:transition}

\begin{figure}[htbp]
	\FigureFile(80mm,80mm){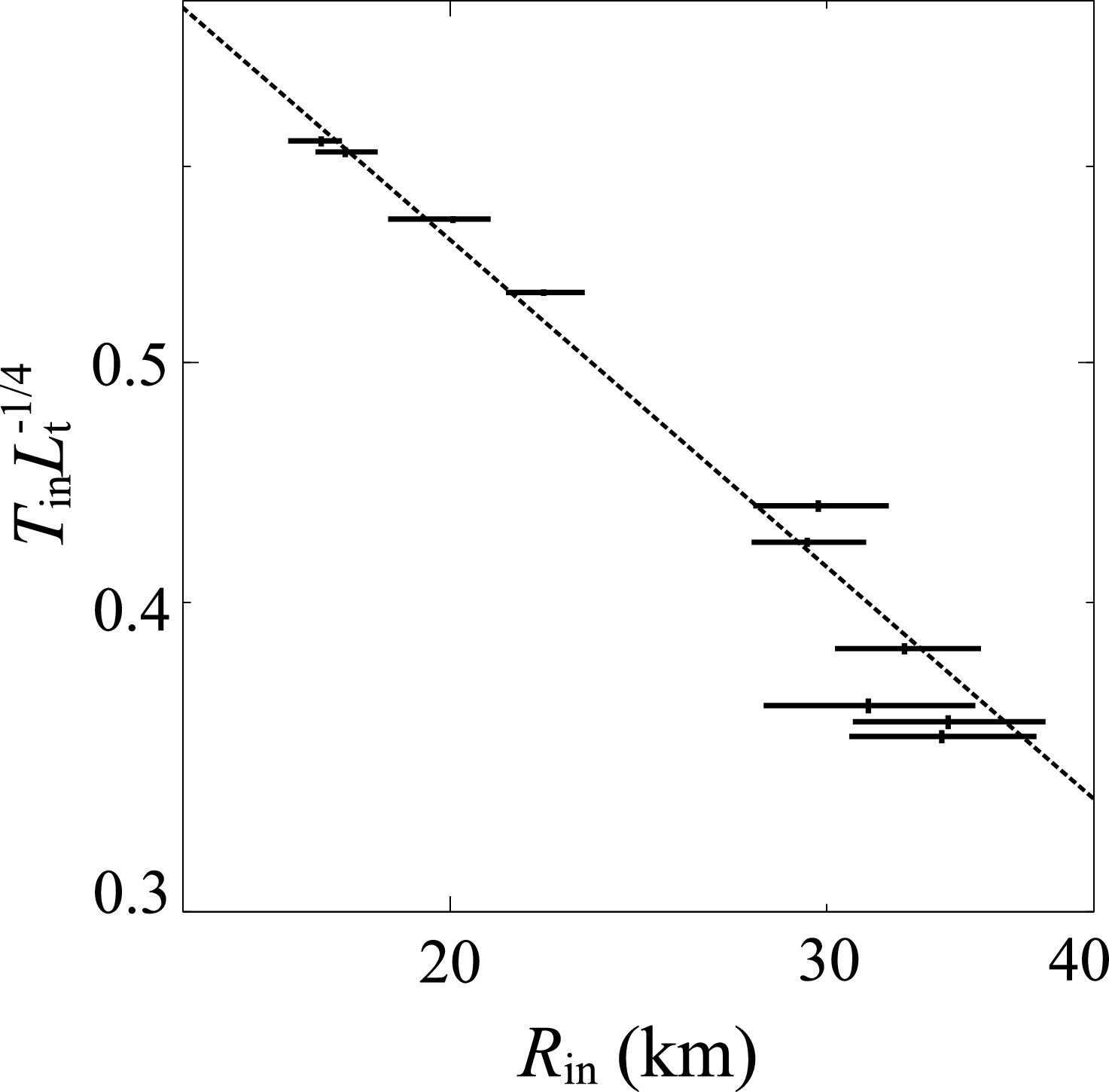}
\caption{A scatter plot between $T_{\rm in}L^{-0.25}_{\rm t}$ and $R_{\rm in}$, covering the complete transition. Dashed line proportional to $R_{\rm in}^{-3/4}$ is  shown as a reference. \label{fig:standard_disk}}
\end{figure}

As shown in figure \ref{fig:parameters_plots}, this precious observation containing the transition has allowed us to directly trace the spectral evolution from the hard to the soft state. As summarized in figure \ref{fig:parameters_plots}, the model parameters changed continuously from typical hard-state values (e.g., \cite{Sakurai14}) to those typical of the soft state of this and similar sources (e.g., \cite{Lin10,Sugizaki13,Sakurai14}). This confirms that the model components in the hard state are identical to their counterparts in the soft state, and provides a support to our interpretation of the less-understood hard state \citep{Sakurai12,Sakurai14,Zhang14,Ono16} in reference to the better-established picture of the soft state (e.g., \cite{Mitsuda84,Takahashi11}).

\par
The transition can be characterized in a straightforward way by the changes of the three luminosities, $L_{\rm BB}$, $L_{\rm comp}$ and $L_{\rm disk}$ (figure \ref{fig:parameters_plots} d). As already pointed out in section \ref{sec:fit1}, these changes can be understood as a process wherein an increased cooling rate causes the outermost part of the coronal flow to progressively change into the innermost region of the optically-thick accretion disk. In fact, $R_{\rm in}$, which is considered to represent the disk--to--corona boundary, decreased from $\sim30$~km to $\sim17$~km, accompanied by a decrease of $T_{\rm e}$ from 10~keV to 3~keV. These changes are also reflected in the decrease of $R_{\rm BB}$ from $\sim12$~km to $\sim6$~km; the accretion in the hard state was occurring rather spherically onto the NS from an inflated corona, whereas that in the soft state became limited to the NS equator as the corona gradually shrank.
\par
We further investigate whether or not the post-transition disk actually continues down to the NS surface. 
In figure \ref{fig:parameters_plots}, $R_{\rm in}$ approached $\sim17$~km (for $i=45^{\circ}$), or $\sim15$~km (for $i=0^{\circ}$). Since these are larger than both $R_{\rm NS}$ and the innermost stable circular orbit (which is also $\sim12$~km for a NS with 1.4 M$_{\odot}$), the disk is likely to be truncated. Because any magnetosphere effect would appear at much lower luminosities ($\sim 10^{36}$~erg s$^{-1}$; \cite{Sakurai14}), the truncation, if true, is likely to be caused by the spontaneous change in the accretion flow, just like in the hard state. However, we are still left with a possibility that the disk actually reaches $R_{\rm NS}$ (or the innermost stable circular orbit), and yet its innermost region (e.g., from $\sim12$~km to $\sim17$~km) is covered by the optically-thick corona to become directly invisible. Thus, it is rather difficult to conclude whether the disk reaches the NS surface or not.
\par

Let us examine our premise that the disk can be regarded as a standard accretion disk. If this assumption is correct, and if the gravitational energy all goes to the X-ray radiation, the P0--P9 parameters should satisfy a scaling as $T_{\rm in} \propto R^{-3/4}_{\rm in}\dot{M}^{1/4}\propto R^{-3/4}_{\rm in}L_{\rm t}^{1/4}$. To see this, in figure \ref{fig:standard_disk} we plot $T_{\rm in}L^{-1/4}_{\rm t}$ against $R_{\rm in}$. Since the expected relation is thus confirmed, we conclude that the disk can be considered standard.

\subsection{Coronal accretion flows}\label{sec:corona}
While the change of the BB and MCD parameters were understood, that of the coronal geometry is yet to be revealed. This can be addressed via the optical depth which represents the amount of the corona along the line of sight. If the coronal geometry were constant across the transition, we would expect $\tau \propto \dot{M} \propto L_{\rm t}$. However, as seen in figure \ref{fig:parameters_plots} (c) and \ref{fig:parameters_plots} (g), $\tau$ remained relatively constant from P0 to P7. This can be explained, at lest qualitatively, by the decrease in $R_{\rm in}$, which made the path length shorter and compensated for the $\dot{M}$ increase. Furthermore, in P8 and P9, $\tau$ increases much more sharply than would be expected by the increase of $\dot{M}$. This may be attributed to two alternative possibilities. One is the decrease in the coronal scale height, as already suggested by \citet{Zhang14}. The other is a decrease in the radial flow velocity $v_{\rm r}(R)$, where $R$ is the distance from the NS center. In either case, the coronal electron density $n_{\rm e}(R)$, averaged over the coronal cross section, will increase, thus making $\tau$ higher.
\par
\begin{figure}[htbp]
	\FigureFile(80mm,80mm){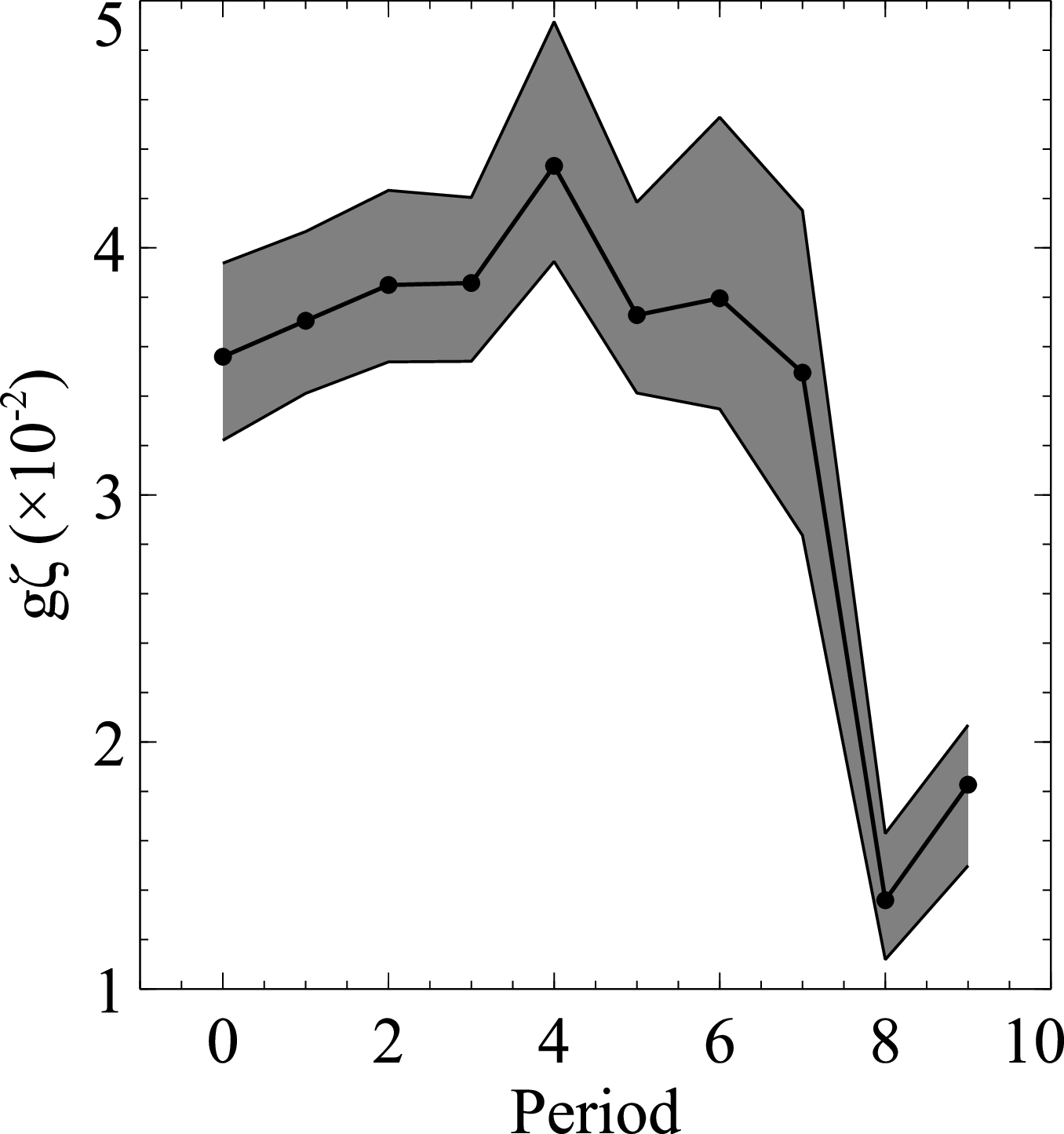}
\caption{The evolution of $g_\zeta$ calculated with equation (\ref{eq:tau2}) and the obtained parameters, plotted as a function of time.\label{fig:g_zeta}}
\end{figure}

After \citet{Sakurai14}, let us quantitatively examine, in several steps, the above two alternatives to explain the final increase in $\tau$. First, in terms of the free-fall velocity $v_{\rm ff}=\sqrt{2G M_{\rm NS}/R}$, we can write as $v_{\rm r}(R)=g v_{\rm ff}(R)$, where $M_{\rm NS}\equiv1.4$~$M_{\rm \odot}$ is the NS mass,  $g$ ($0<g<1$) is a numerical factor which is assumed to be a constant over the coronal volume (but can vary through the transition). Next, $\tau$ can be expressed as \par
\begin{equation}\label{eq:tau1}
\tau=\sigma_{\rm T} \int n_{\rm e} dl
\end{equation}
where $\sigma_{\rm T}$ is the Thomson scattering cross section, and the integration is made along our line of sight. Third, $v_{\rm r}(R)$ and $n_{\rm e}$ can be incorporated together into the mass accretion rate $\dot{M}$ as\par
\begin{equation}\label{eq:Mdot}
\dot{M}=S(R)v_{\rm r}(R)\mu m_{\rm p}n_{\rm e}(R)
\end{equation}
where $S(R)$ is the coronal cross section at $R$, $\mu \sim 1.2$ is the average molecular weight and $m_{\rm p}$ is the proton mass. In addition, we may assume, for simplicity, that $S(R)$ is approximated as $S(R)=4\pi R^2 \zeta$, where $\zeta$ ($0<\zeta \leq1$) is a form factor. Then, as long as $\zeta$ is rather close to unity, the integration in equation (\ref{eq:tau1}) can be approximated by radial integration from $R=R_{\rm NS}$ to $R_{\rm in}$. Finally, performing this integration, and eliminating $\dot{M}$ through the relation $L_{\rm t}=GM_{\rm NS} \dot{M}/R_{\rm NS}$, we obtain\par
\begin{equation}\label{eq:tau2}
\tau=0.16\times \frac{1}{g \zeta}\times \left(\frac{1.2}{\mu}\right) \times \left(\sqrt{\frac{R_{\rm in}}{R_{\rm NS}}}-1\right)\times \frac{L_{\rm t} ({\rm erg~s^{-1}})}{10^{37} ({\rm erg~s^{-1}})}
\end{equation}
Equating equation (\ref{eq:tau2}) with the observed $\tau$, and substituting the observed values of $R_{\rm in}$ and $L_{\rm t}$, we have deduced the behavior of $g\zeta$ as shown in figure \ref{fig:g_zeta}. Thus, the factor $g\zeta$ over P0--P7 has been obtained as $g=0.035-0.043$, in agreement with the value of $g=0.04$ (assuming $\zeta=1$) derived by \citet{Sakurai14} in the hard state. Furthermore, it steeply decreases to $g\zeta=0.014-0.018$ at P8 and P9.

\par
In this way, the observed increase in $\tau$ towards the final stage of the transition can be explained by a decrease in either $\zeta$, or $g$, or both. The former means the flattening of the corona as already discussed, while the latter means a decrease of the radial flow velocity. In any case, an important consequence is the requirement for $g\ll1$ (assuming $\zeta$ is not too small), as pointed out by \citet{Sakurai14}; that is, the radial flow velocity of the corona is much lower than the free-fall velocity, and hence the corona is still in a nearly Keplerian motion (particularly at later phases of the transition) with the azimuthal velocity component dominating over $v_{\rm r}$.
\par
We implicitly assumed so far that the density of the corona, by definition, is low enough not to emit photons by itself. In order to confirm this, the bremsstrahlung emissivity of the corona was estimated from the electron density in equation (\ref{eq:Mdot}) and utilizing $g \zeta=0.013-0.043$. The bremsstrahlung emissivity, calculated as $\int n_{\rm e}^2 dV \sim n^2_{e}(g \zeta,R_{\rm in})\times(4\pi/3) \times R_{\rm in}^3$, then becomes $< 8.6\times 10^{56}$~cm$^{-3}$ over P0--P9. This corresponds to a luminosity of $1.3\times10^{34}$~erg s$^{-1}$ \citep{Chakrabarty14}, which is negligible compared with $L_{\rm comp}>5\times10^{36}$~erg s$^{-1}$. Thus, our scenario is self consistent.

\subsection{Energy budget}\label{sec:budget}
In figure \ref{fig:parameters_plots} (d), we find $L_{\rm BB}\simeq0.3L_{\rm t}$ throughout. Although the consistency of the $L_{\rm BB}/L_{\rm t}$ ratio is consistent with the prediction by the virial theorem applied to the disk and the corona in a Keplerian motion, the absolute value of $\sim0.3$ is lower than the expected value of 0.5. An obvious cause of this discrepancy is in the system inclination; it may be lower than the assumed value of 45$^{\circ}$. Another interesting possibility is soft landing of the accreting gas onto the NS surface. The NS in this system is considered to be spinning with the frequency $f=550.27$~Hz \citep{Casella08}. Thus, assuming that the accreting gas is rotating with the Keplerian velocity $v_{\rm K}$ at $R=R_{\rm NS}$ (as evidenced by the low value of $g$), the BB luminosity would be reduced by a factor of $1-(2\pi f R_{\rm NS}/v_{\rm K})^2\simeq0.88$. Yet another possibility is azimuthal bulk Comptonization of the BB photons by the high circular velocity of the corona (section \ref{sec:corona}); this will take up a fraction of the Keplerian energy (up to $L_{\rm comp}/L_{\rm t}\sim11\%$) that should have been radiated as $L_{\rm BB}$. Considering these multiple effects, the result in figure \ref{fig:parameters_plots} (d) may be brought into a better agreement with the virial theorem.
\par
In addition to the above two possibilities, yet another obvious explanation to the 20\% deficit of $L_{\rm BB}$ is to assume that a fraction of matter is launched as outflows instead of accreting onto the NS surface (e.g., \cite{Fukue04}). According to \citet{Takahashi11}, 4U1608-52 indeed showed a sign of such outflows when its total luminosity was $1-4\times10^{37}$~erg s$^{-1}$. However, in contrast to the case of 4U 1608-52 wherein the $L_{\rm BB}/L_{\rm tot}$ ratio decreased meanwhile from $\sim 0.5$ to $\sim 0.25$, the ratio in Aql X-1 remained relatively constant at $\sim0.3$ as the luminosity increases during the transition. We hence consider that outflows would not provide a major source of the $L_{\rm BB}$ deficit in the present case.

\subsection{Transition Time Scales}\label{sec:timescale}
Based on continuous monitoring of LMXBs with MAXI and Swift, \citet{Asai15} showed that the state transitions in LMXBs occur in two types; one type of events, involved in ``normal outbursts'', take place on a typical time scale of a day or less, while those of the other type, seen in ``mini-outbursts'' with much smaller flux changes, proceed on a longer time scales of a few days. As seen in the MAXI light curve (figure 3 of \cite{Asai15}), the present transition is clearly associated with a normal outburst in their classification, and the transition time scale we observed, about 30 ks, or $\sim 8$~hours, is consistent with those seen in that type of transitions.
\par
Compared with these episodes in LMXB, the soft vs. hard state transitions in black-hole binaries have been observed to take place on considerably longer time scales, several days to 10 days (e.g., \cite{Maejima84,Remillard06,Nakahira12}). A simple explanation of this difference would be that the radial propagation of changes in the accretion-disk conditions (typically on viscous time scales) takes longer in stellar black holes due to the larger disk size reflecting the mass difference. Alternatively, the difference may be attributed to the stronger Compton cooling of a corona in an LMXB, by the BB photons which are absent in black holes. Further investigation into this issue is beyond the scope of the present paper.

\section{Conclusion}
We analyzed an archival data of Aql X-1 taken with Suzaku on 2011 October 21 for a gross duration of 80 ks. The source underwent a transition from the hard state to the soft state, on a time scale of $\sim 30$~ks. We have successfully explained the spectral evolution by combining the disk blackbody emission with thermal Comptonization of the blackbody photons from the NS surface. The model parameters changed continuously, which justifies the spectral decomposition, and unifies the interpretations of the soft and the hard state. Utilizing the obtained parameters, particularly $\tau$, the Comptonizing corona is suggested to shrink radially as the source approaches the soft state. In addition, the transition is likely to involve a vertical flattening of the corona, or a decrease in the radial flow velocity in the corona, or both.

\begin{ack}
This work was supported by a Grant-in-Aid for JSPS Fellows No. 15J08913
\end{ack}




\end{document}